\documentclass[dvips]{article}
\usepackage{amssymb}
\usepackage{amsmath}
\usepackage{rotating}
\DeclareSymbolFont{ppa}{OT1}{ppl}{m}{it}
\DeclareMathSymbol{\vv}{\mathalpha}{ppa}{'166}

\begin{document}

\newcommand{\dd}{\,{\rm d}}
\newcommand{\ie}{{\it i.e.},\,}
\newcommand{\etal}{{\it et al.\ }}
\newcommand{\eg}{{\it e.g.},\,}
\newcommand{\cf}{{\it cf.\ }}
\newcommand{\vs}{{\it vs.\ }}
\newcommand{\zdot}{\makebox[0pt][l]{.}}
\newcommand{\up}[1]{\ifmmode^{\rm #1}\else$^{\rm #1}$\fi}
\newcommand{\dn}[1]{\ifmmode_{\rm #1}\else$_{\rm #1}$\fi}
\newcommand{\upd}{\up{d}}
\newcommand{\uph}{\up{h}}
\newcommand{\upm}{\up{m}}
\newcommand{\ups}{\up{s}}
\newcommand{\arcd}{\ifmmode^{\circ}\else$^{\circ}$\fi}
\newcommand{\arcm}{\ifmmode{'}\else$'$\fi}
\newcommand{\arcs}{\ifmmode{''}\else$''$\fi}
\newcommand{\MS}{{\rm M}\ifmmode_{\odot}\else$_{\odot}$\fi}
\newcommand{\RS}{{\rm R}\ifmmode_{\odot}\else$_{\odot}$\fi}
\newcommand{\LS}{{\rm L}\ifmmode_{\odot}\else$_{\odot}$\fi}
\newcommand{\feh}{\hbox{$ [{\rm Fe}/{\rm H}]$}}

\newcommand{\Abstract}[2]{{\footnotesize\begin{center}ABSTRACT\end{center}
\vspace{1mm}\par#1\par
\noindent
{~}{\it #2}}}

\newcommand{\TabCap}[2]{\begin{center}\parbox[t]{#1}{\begin{center}
  \small {\spaceskip 2pt plus 1pt minus 1pt T a b l e}
  \refstepcounter{table}\thetable \\[2mm]
  \footnotesize #2 \end{center}}\end{center}}

\newcommand{\TableSep}[2]{\begin{table}[p]\vspace{#1}
\TabCap{#2}\end{table}}

\newcommand{\FigCap}[1]{\footnotesize\par\noindent Fig.\  %
  \refstepcounter{figure}\thefigure. #1\par}

\newcommand{\TableFont}{\footnotesize}
\newcommand{\TableFontIt}{\ttit}
\newcommand{\SetTableFont}[1]{\renewcommand{\TableFont}{#1}}

\newcommand{\MakeTable}[4]{\begin{table}[htb]\TabCap{#2}{#3}
  \begin{center} \TableFont \begin{tabular}{#1} #4
  \end{tabular}\end{center}\end{table}}

\newcommand{\MakeTableSep}[4]{\begin{table}[p]\TabCap{#2}{#3}
  \begin{center} \TableFont \begin{tabular}{#1} #4
  \end{tabular}\end{center}\end{table}}

\newenvironment{references}%
{
\footnotesize \frenchspacing
\renewcommand{\thesection}{}
\renewcommand{\in}{{\rm in }}
\renewcommand{\AA}{Astron.\ Astrophys.}
\newcommand{\AAS}{Astron.~Astrophys.~Suppl.~Ser.}
\newcommand{\ApJ}{Astrophys.\ J.}
\newcommand{\ApJS}{Astrophys.\ J.~Suppl.~Ser.}
\newcommand{\ApJL}{Astrophys.\ J.~Letters}
\newcommand{\AJ}{Astron.\ J.}
\newcommand{\IBVS}{IBVS}
\newcommand{\PASJ}{PASJ}
\newcommand{\PASP}{P.A.S.P.}
\newcommand{\Acta}{Acta Astron.}
\newcommand{\MNRAS}{MNRAS}
\renewcommand{\and}{{\rm and }}
\section{{\rm REFERENCES}}
\sloppy \hyphenpenalty10000
\begin{list}{}{\leftmargin1cm\listparindent-1cm
\itemindent\listparindent\parsep0pt\itemsep0pt}}%
{\end{list}\vspace{2mm}}

\def\TYLDA{~}
\newlength{\DW}
\settowidth{\DW}{0}
\newcommand{\dw}{\hspace{\DW}}

\newcommand{\refitem}[5]{\item[]{#1} #2%
\def\REFARG{#3}\ifx\REFARG\TYLDA\else, {\it#3}\fi
\def\REFARG{#4}\ifx\REFARG\TYLDA\else, {\bf#4}\fi
\def\REFARG{#5}\ifx\REFARG\TYLDA\else, {#5}\fi.}

\newcommand{\Section}[1]{\section{#1}}
\newcommand{\Subsection}[1]{\subsection{#1}}
\newcommand{\Acknow}[1]{\par\vspace{5mm}{\bf Acknowledgments.} #1}
\pagestyle{myheadings}

\newfont{\bb}{ptmbi8t at 12pt}
                      \newcommand{\xrule}{\rule{0pt}{2.5ex}}
\newcommand{\xxrule}{\rule[-1.8ex]{0pt}{4.5ex}}
\def\thefootnote{\fnsymbol{footnote}}

\begin{center}
{\Large\bf Properties of the Milky Way's Old Populations Based on
Photometric Metallicities of the OGLE RR Lyrae Stars}
\vskip1cm
{\bf
P.~~P~i~e~t~r~u~k~o~w~i~c~z$^1$,~~A.~U~d~a~l~s~k~i$^1$,~~I.~~S~o~s~z~y~\'n~s~k~i$^1$,\\
D.~M.~~S~k~o~w~r~o~n$^1$,~~M.~~W~r~o~n~a$^1$,~~M.~K.~~S~z~y~m~a~\'n~s~k~i$^1$,
R.~~P~o~l~e~s~k~i$^{1}$,~~K.~~U~l~a~c~z~y~k$^{1,2}$,~~S.~~K~o~z~{\l}~o~w~s~k~i$^1$,
J.~~S~k~o~w~r~o~n$^1$,~~P.~~M~r~\'o~z$^{1,3}$,~~K.~~R~y~b~i~c~k~i$^1$,\\
P.~~I~w~a~n~e~k$^1$,~~and~~M.~~G~r~o~m~a~d~z~k~i$^1$\\}
\vskip3mm
{
$^1$ Astronomical Observatory, University of Warsaw, Al. Ujazdowskie 4, 00-478 Warszawa, Poland\\
e-mail: pietruk@astrouw.edu.pl\\
$^2$ Department of Physics, University of Warwick, Coventry CV4 7AL, UK\\
$^3$ Division of Physics, Mathematics, and Astronomy, California Institute of Technology, Pasadena, CA 91125, USA\\
}
\end{center}

\Abstract{
We have used photometric data on almost 91~000 fundamental-mode RR Lyrae stars
(type RRab) detected by the OGLE survey to investigate properties of old populations
in the Milky Way. Based on their metallicity distributions, we demonstrate that
the Galaxy is built from three distinct old components: halo, bulge, and disk.
The distributions reach their maxima at approximately ${\rm [Fe/H]_{J95}}=-1.2$,
$-1.0$, and $-0.6$ dex on the Jurcsik's metallicity scale, respectively.
We find that, very likely, the entire halo is formed from infalling dwarf
galaxies. It is evident that halo stars penetrate the inner regions
of the Galactic bulge. We estimate that about one-third of all RR Lyr stars
within the bulge area belong in fact to the halo population. The whole old bulge
is dominated by two populations, A and B, represented by a double sequence in the
period--amplitude (Bailey) diagram. The boundary in iron abundance between
the halo and the disk population is at about ${\rm [Fe/H]_{J95}}=-0.8$ dex.
Using {\it Gaia} DR2 for RRab stars in the disk area, we show that the observed
dispersion of proper motions along the Galactic latitude decreases smoothly with
the increasing metal content excluding a bump around ${\rm [Fe/H]_{J95}}=-1.0$ dex.
}

{Galaxy: bulge -- Galaxy: disk -- Galaxy: halo -- Stars: variables: RR Lyrae}


\Section{Introduction}

Variable stars of RR Lyr type are pulsating yellow giants on the
horizontal branch (Feuchtinger 1999, Marconi \etal 2015). Thanks to their
relatively high luminosity (of about 50~$\LS$) and characteristic
high-amplitude light curves, the variables are detected within the whole Local
Group (\eg Mart\'inez-V\'azquez \etal 2016, 2017; Cusano \etal 2017). RR Lyr stars
serve as standard candles and tracers of old populations (older than 10 Gyr).
Thousands of variables of this type have been found in the Galactic bulge
(\eg Soszy\'nski \etal 2014), Galactic halo (\eg Torrealba \etal 2015),
and halos of Local Group galaxies (\eg Sarajedini \etal 2012). The variables
have been used in the determination of the distance to the Galactic center
(\eg Pietrukowicz \etal 2015, Majaess \etal 2018) and numerous Galactic globular
clusters (\eg Braga \etal 2015, 2016; Tsapras \etal 2017), also in the studies
of the structure of the Milky Way (\eg Alcock \etal 1998, Pietrukowicz \etal 2012,
D\'ek\'any \etal 2013, Zinn \etal 2014, Prudil \etal 2019) and
Magellanic Clouds (Jacyszyn-Dobrzeniecka \etal 2017).
RR Lyr stars as old objects provide important hints to the
understanding of the formation and early evolution of the Milky Way
(\eg Kinemuchi \etal 2006, Szczygie{\l} \etal 2009,
Fiorentino \etal 2015, 2017, Belokurov \etal 2018).

Amongst the known RR Lyr variables there are fundamental-mode pulsators
(RRab stars), first-overtone pulsators (RRc stars), and rarely observed
classical and anomalous multiple-mode pulsators (RRd stars).
RRab stars have several useful advantages over the other types.
Variables of the RRab type have characteristic saw-tooth-shaped phase-folded
light curves in the optical range. This significantly facilitates their
proper classification in contrast to nearly sinusoidal RRc stars. Amplitudes
of RRab stars are generally much higher than those of the other types.
RRab stars get brighter than RRc stars with the increasing wavelength.
These properties make the searches for RRab stars highly complete
in the optical bands, in particular in the $I$ band. At longer
wavelengths, in near-infrared and mid-infrared, the amplitudes
of the variables are smaller and the light curve shapes are more rounded,
which hampers the detection and often leads to misclassification.

Finally, there is a very practical property of RRab stars that we use here.
It is possible to assess metallicity $\feh$ of the star based on its
pulsation period and light curve shape. The method of estimating
photometric metallicity was proposed by Kov\'acs and Zsoldos (1995)
and developed by Jurcsik and Kov\'acs (1996).

In this work, we exploit the largest available homogeneous collection of
RR Lyr stars, part of the OGLE Collection of Variable
Stars\footnote{http://www.astrouw.edu.pl/ogle/ogle4/OCVS/},
to describe the Milky Way's ancient
populations based on the information on photometric metallicity. 
The Optical Gravitational Lensing Experiment (OGLE) is a long-term
variability survey conducted on the 1.3-m Warsaw telescope at the
Las Campanas Observatory, Chile, of the Carnegie Institution for Science.
Regular, high-quality OGLE observations (Udalski \etal 2015) of the Galactic
bulge, Galactic disk, Small and Large Magellanic Clouds (SMC and LMC) have
allowed the detection and classification of over 126~000 genuine RR Lyr stars
amongst about one million variable objects of various types. Light curves
of all OGLE periodic variables are inspected visually, which
makes the collection pure and highly complete. The photometry
is extracted with the help of image subtraction technique,
especially suitable for crowded stellar fields (Wo\'zniak 2000).
Published time-series data contain hundreds to thousands of $I$-band
measurements per star over at least two observing seasons. This
guarantees well-sampled phase-folded light curves of RR Lyr variables, accurate
determination of their pulsation periods, and the detection of possible
amplitude and phase modulation, namely the presence of the Blazhko effect.

The latest releases of RR Lyr stars detected in the Magellanic System area
and in the stripe of the Milky Way contain 47~828 and 78~350 objects and were
published by Soszy\'nski \etal (2019a) and Soszy\'nski \etal (2019b), respectively.
The analysis presented below is based on 56~424 RRab stars from the Galactic
bulge/disk fields covering an area of 2750 square degrees and 34~229 RRab
stars from the Magellanic System fields covering 765 square degrees.


\Section{Preparation of the samples}

Proper analysis of the Milky Way's populations requires cleaning the RRab
collection from members of Galactic globular clusters and members of nearby
irregular galaxies. To clean our RRab samples from globular cluster
stars, we used the updated catalog of variable stars in globular
clusters\footnote{http://www.astro.utoronto.ca/$\sim$cclement/read.html}
published by Clement \etal (2001). We cross-matched the catalog with the
OGLE collection, verified each star individually, and noted only
{\it bona fide} members and very likely members of the globular clusters.
As a result, from the entire list of OGLE RRab stars observed in the
Galactic stripe we removed 369 stars residing in 22 (mostly bulge)
globular clusters. From the set of RRab stars observed in the
Magellanic System area, we removed 18 variables from Galactic globular
clusters 47 Tucanae and NGC 362 located in the foreground of the SMC.

To separate Galactic bulge from Sagittarius dwarf spheroidal (Sgr dSph)
galaxy and to extract a Galactic halo sample from the background Magellanic
Clouds, we applied simple brightness criteria. In Fig.~1, we present
distributions of the mean $I$-band magnitudes of RRab variables from
the entire Magellanic System area and a part of the Galactic bulge area
containing the Sgr dSph core, where the applied magnitude boundaries
are marked with vertical lines. The bulge part was cropped to
$+10\arcd>l>0\arcd$ and $b<-5\arcd$ after a series of experiments to check
how far from the center of Sgr dSph the variables from this galaxy are detected.
We found that, for instance, in the bulge area located on the opposite
side of the Galactic center, for $l<0\arcd$ and $b>+5\arcd$, there is no
excess of detections around $I=17.5$ mag in the brightness distribution.
This magnitude limit was applied to variables detected in the shallow
bulge survey, for which there is no $V$-band data so far. For variables
from the regular OGLE-IV bulge survey (Soszy\'nski \etal 2014) we used
information on the $V-I$ color. We made exactly the same cuts in the
color-magnitude diagram as shown in Pietrukowicz \etal (2015).

\begin{figure}[htb!]
\centerline{\includegraphics[angle=0,width=90mm]{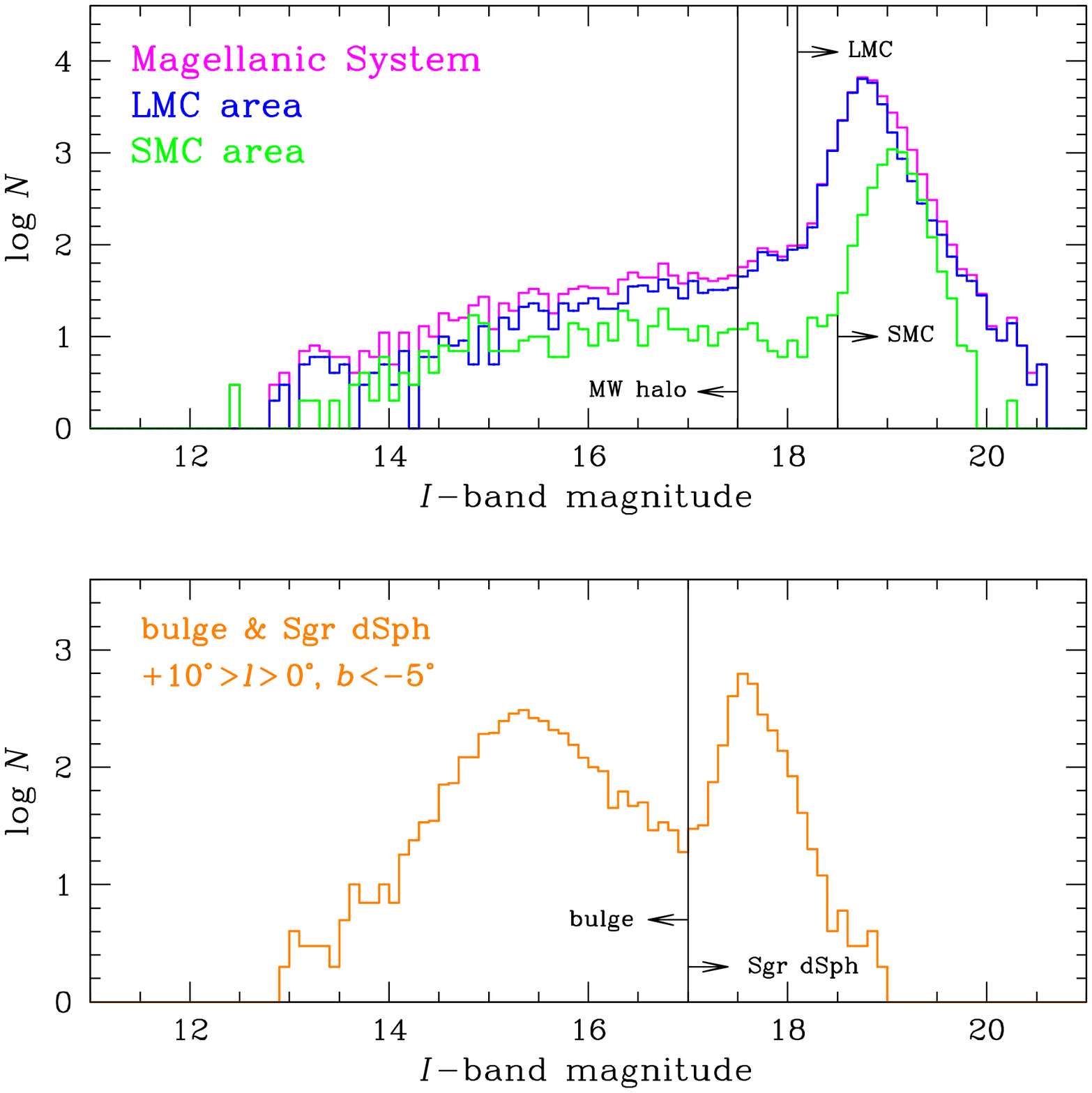}}
\FigCap{
Brightness distribution of OGLE RRab stars observed toward the Magellanic
System (upper panel) and a selected bulge area with Sgr dSph galaxy in the
background (lower panel). The following magnitude limits are used
to select members of the nearby galaxies: $I=18.1$ for the LMC,
$I=18.5$ for the SMC, and $I=17.0$ for Sgr dSph. RRab stars
observed toward the Magellanic System that are brighter than $I=17.5$ mag
belong to the Milky Way's halo.
}
\end{figure}

As a result of the above operations, we extracted the following samples
of RRab variables representing nearby galaxies: 2288 stars from the
Sgr dSph, 26~947 stars from the LMC, and 4601 stars from the SMC. By fitting
a straight line (power-law) to the brightness distribution on the far side
of the bulge (for $15.5>I>17$ mag) and extrapolating it to fainter magnitudes,
we estimate the contamination of the Sgr dSph sample by Milky Way stars at
a level of 7\%, or its purity of about 93\%. The purity of the samples
of more distant Magellanic Clouds is of about 97\% assuming a constant
level of contamination by halo stars in the brightness distribution.
An estimated number of Sgr dSph variables left in the bulge sample is about
30. This is a negligible value for our analysis. To obtain a pure Galactic halo
sample, we additionally removed RRab variables with $I$-band amplitudes
below 0.2 mag located within a radius of $4\arcd$ from the LMC center.
Those are real but blended LMC objects, as it was noticed by
Jacyszyn-Dobrzeniecka \etal (2017).

After the correction for globular cluster and Sgr dSph objects our final set
of Galactic bulge and disk RRab stars contains 52~994 variables. The final halo
sample contains 951 RRab stars. In total, 53~945 stars remained as representatives
of the Milky Way's populations. Upper panels of Fig.~2 present the bulge area before
and after the cleaning operation. Note that the area around the Sgr dSph
center at Galactic coordinates ($l$,$b$)=($+6\arcd$,$-14\arcd$) seems to be
free from stars from that galaxy. Lower panel of Fig.~2 shows the map of all RRab
stars prepared for further analysis.

\begin{figure}[h!]
\centerline{\includegraphics[angle=0,width=124mm]{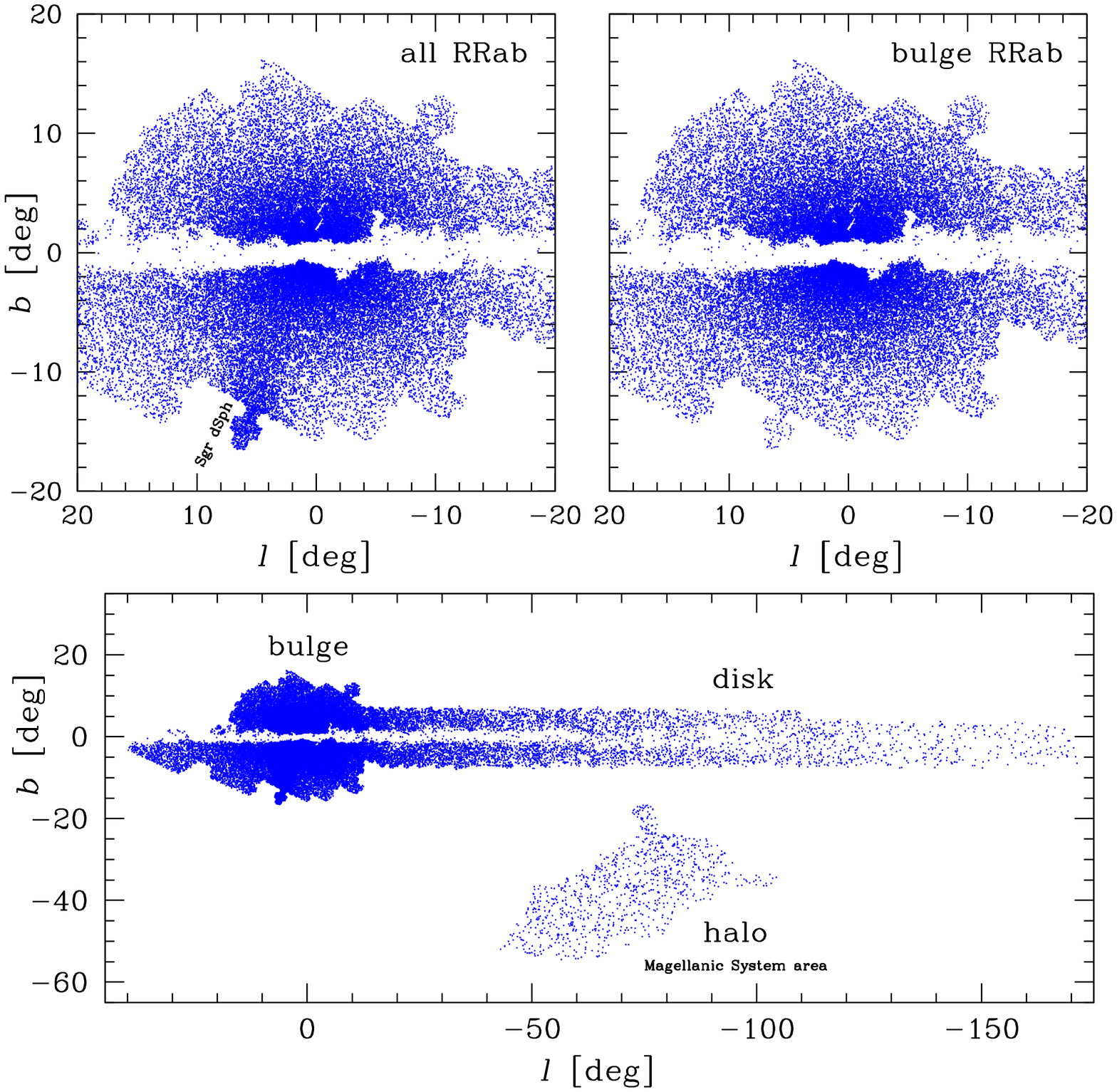}}
\FigCap{
Upper left panel: distribution of OGLE RRab variables detected
toward the Galactic bulge. Upper right panel: the sample cleaned
for globular cluster members and stars from Sgr dSph galaxy.
Lower panel: distribution of 53~945 only-Galactic RRab variables observed
in all OGLE fields covering about 3515 square degrees of the sky.
Due to extreme extinction along the Galactic plane, many stars
in this region are not visible in the optical range.
}
\end{figure}


\Section{Metallicity Distributions}

We estimate metallicities of the OGLE RRab stars using the following formula
from Smolec (2005):
\begin{equation}
{\rm \feh_{J95}} = - 3.142 - 4.902 P + 0.824 \phi_{31},
\end{equation}
where $P$ is the pulsation period and $\phi_{31}=\phi_3-3\phi_1$
is a Fourier phase combination for sine decomposition of the $I$-band
light curve. The obtained metallicities are on the Jurcsik's (1995)
scale (J95). The above formula was calibrated based on spectroscopic
and photometric data for 28 RRab stars with iron abundances ranging
from $-1.7$ to $+0.1$ dex and located in the Galactic field and two
globular clusters. The formula provides a straightforward calculation of the
metallicity using the $I$-band photometry. A weak point of this relation
is its linearity and the lack of metal-poor stars for calibration.
It is known to overestimate the iron abundance at the metal-poor end.
According to Smolec (2005), the uncertainty of the relation is 0.18~dex.

We note that a more accurate, non-linear formula to calculate photometric
metallicities was later proposed by Nemec \etal (2013) based on 34 RRab
variables located in the original field of view of the {\it Kepler} space
mission. The range of used metallicities is wider and spans from $-2.6$
to 0.0 dex. However, the formula from Nemec \etal (2013) in the application
to the rich $I$-band OGLE measurements would require the use of additional
transformations of the Fourier phase combination $\phi_{31}$ from the
$I$-band to the $V$-band and from the $V$-band to the {\it Kepler} filter,
provided for instance by Skowron \etal (2016) and Jeon \etal (2014),
respectively. We found that these transformations generate a significant
scatter making the metallicity distribution wider than expected on the
metal-rich side. This is also in part due to the fact that the relation
of Nemec \etal (2013) is calibrated only up to $\feh=0.0$ dex, which,
combined with its non-linear dependence on $\phi_{31}$, may overestimate
the photometric metallicity at the metal-rich end. Since the majority
of bulge RRab fall within a calibration region of the relation of
Smolec (2005), we decide to use this relation in our study.

In Fig.~3, we present metallicity distributions for various sections
of the Milky Way and for nearby irregular galaxies. In the upper panel of
this figure, we compare distributions for inner bulge (stars observed
within the angular radius $r=10\arcd$ from the origin of the Galactic
coordinates), outer bulge (stars with $10\arcd<r\leq20\arcd$),
bulge-to-disk transition area ($20\arcd<r\leq30\arcd$),
disk area ($r>30\arcd$), and for the Galactic halo sample extracted
from the Magellanic System fields. Both bulge distributions show
a sharp peak at ${\rm [Fe/H]_{J95}}\approx-1.0$ dex. This peak is less
prominent in the bulge-to-disk transition region. In the disk area,
there are two maxima at different iron abundance, a larger one at
${\rm [Fe/H]_{J95}}\approx-1.2$ dex and a smaller one at
${\rm [Fe/H]_{J95}}=-0.6$ dex. The shape of the pure halo population
allows us to explain the mentioned distributions.
The metal-poor sides of all distributions are very similar
to each other---they have the convex shape of the halo distribution.
This means that the halo population is present in all parts of the Milky Way
including its inner regions. The larger of the two maxima in the disk area
(at ${\rm [Fe/H]_{J95}}\approx-1.2$ dex) corresponds to the halo distribution.
The metal-rich component certainly belongs to the old disk population
as this component is not seen at high Galactic latitudes.

\begin{figure}[h!]
\centerline{\includegraphics[angle=0,width=124mm]{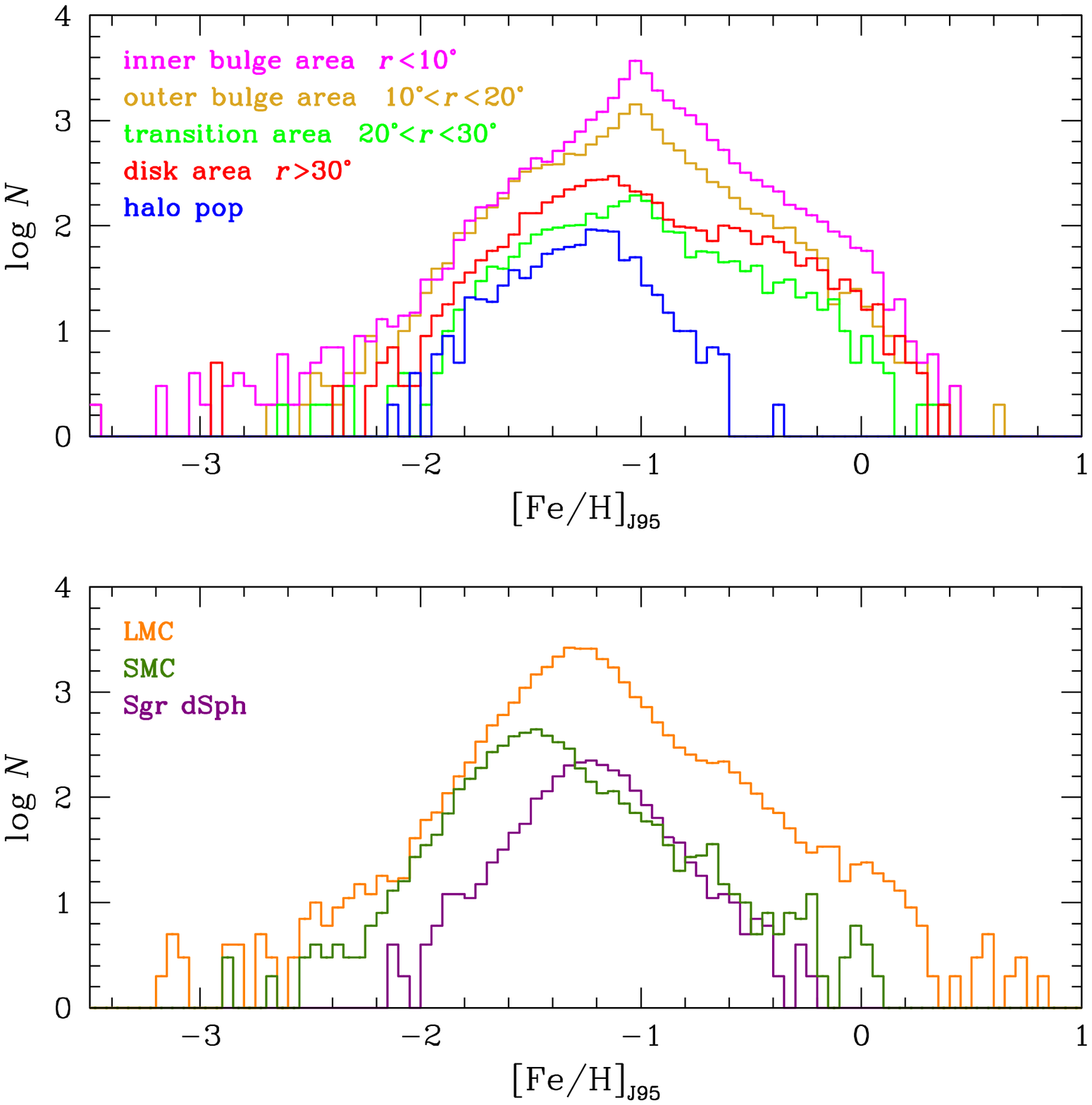}}
\FigCap{
Metallicity distributions for RRab stars observed in various directions
in the Milky Way (upper panel) and in nearby galaxies (lower panel).
Note a very similar, convex shape of the metal-poor side of the
distributions in all directions indicating that the halo penetrates
down to the very central regions of the Galactic bulge.
The distribution for the disk area has two maxima, a metal-poor one
corresponding to the Galactic halo population and a metal-rich
one corresponding to the old (thick) disk population.
}
\end{figure}

We note here that foreground disk stars contribute insignificantly to our
bulge sample. There is no sharp boarder between the disk and the bulge.
However, we can safely assume that RR Lyr stars observed toward the Galactic
bulge and located closer than 3 kpc from the Sun do not belong to the
bulge. Approximately, at that distance we can find globular cluster M22
($3.1\pm0.2$ kpc, Kunder \etal 2013). Only 55 RRab variables from
the bulge area have the mean brightness $I<13.1$ mag or are brighter
than RR Lyr stars from M22. This low number of bright stars in the
sample also stems from the saturation limit in the OGLE regular
bulge survey of $I\approx12.5$ mag. The mentioned 55 RRab stars may
belong to the disk as well as to the halo population. Therefore,
we decided not to apply any cuts for bright variables from the bulge
fields in the cleaning process.

In Fig.~4, we try to estimate what fraction of halo RR Lyr stars is present
in the bulge. We scaled the halo profile and subtracted it from the
bulge profile. The scaling factor was found by minimizing the difference
between the two profiles for ${\rm [Fe/H]_{J95}}\leq-1.4$ dex. It turns out
that within $r=20\arcd$ from the center about 32\% of the RR Lyr stars
are halo interlopers. For $r<10\arcd$ and $r<5\arcd$ we obtained 25\% and 19\%,
respectively. This is consistent with a very recent result
by Kunder \etal (2020). Their spectroscopic survey of 2768 RRab variables
in the area $+8\arcd>l>-8\arcd$ and $-6\arcd<b<-3\arcd$ showed that
25\% of the sample stars belongs to the halo population.

\begin{figure}[htb!]
\centerline{\includegraphics[angle=0,width=124mm]{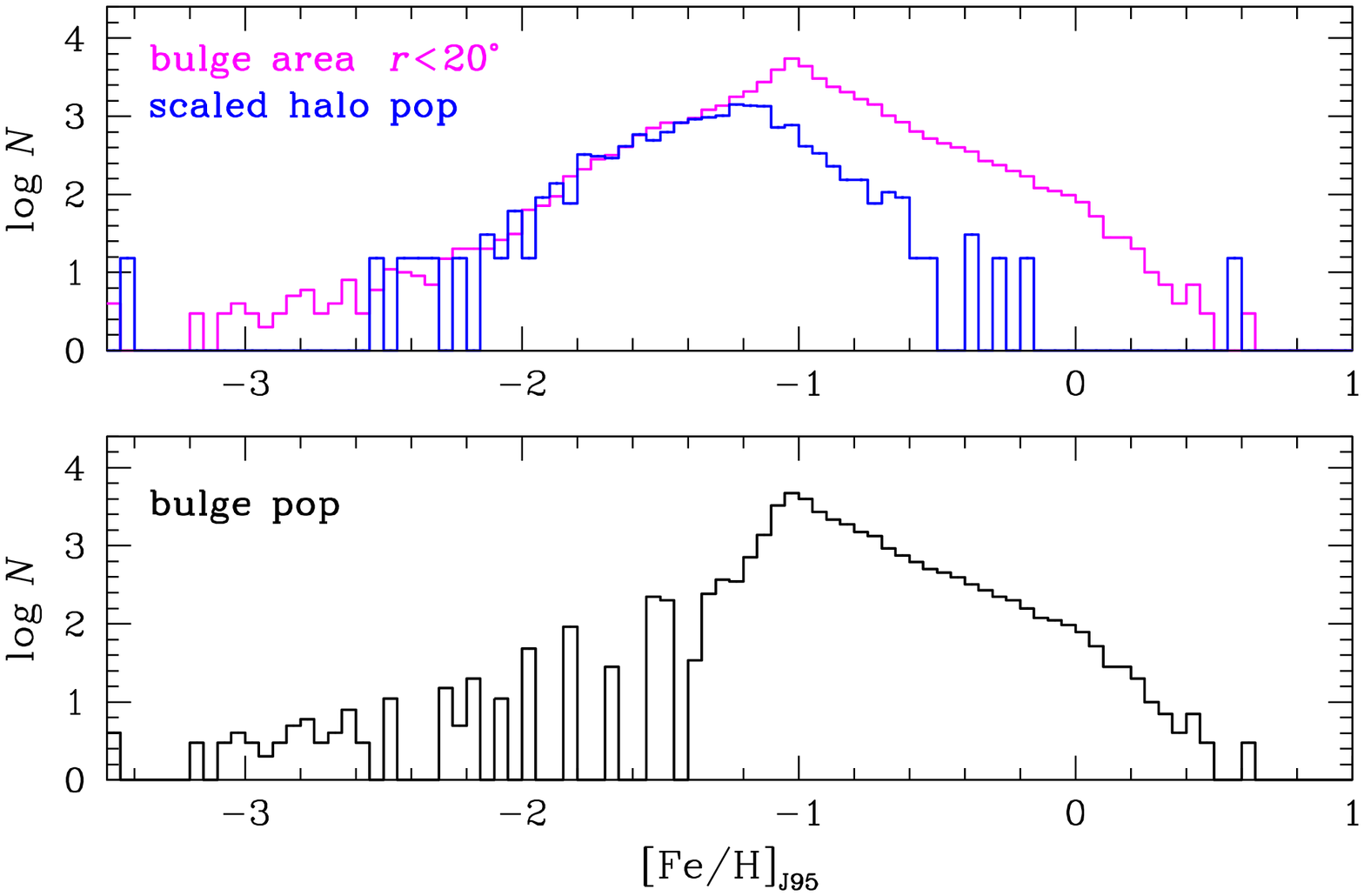}}
\FigCap{
Extraction of the metallicity profile for the bulge population
by subtraction of the halo profile scaled to the bulge profile.
In the whole bulge area, about 32\% of RR Lyr stars are halo interlopers.
}
\end{figure}

In the lower panel of Fig.~3, we present metallicity histograms for the
three nearby irregular galaxies observed in the OGLE project: LMC, SMC,
and Sgr dSph. For both Magellanic Clouds the OGLE collection is
complete, while the disrupted Sgr dSph galaxy has only its core covered.
The metallicity distributions for LMC, SMC, and Sgr dSph reach
their maxima at about $-1.3$, $-1.5$, and $-1.2$ dex, respectively.

In the top three panels of Fig. 5, we compare normalized distributions for the
nearby galaxies with the distribution for the Galactic halo. Sgr dSph seems
to have the most similar metallicity distribution to the one of the halo. In the
bottom panel of Fig. 5, we present the result of a simple experiment.
We mixed RR Lyr samples from the three galaxies in equal proportion and
normalized the obtained mixture to the halo. Such artificial population has an
even more similar profile to the one of the halo. This experiment shows that
the entire Galactic halo might be composed of stars from the accreted galaxies.

\begin{figure}[htb!]
\centerline{\includegraphics[angle=0,width=124mm]{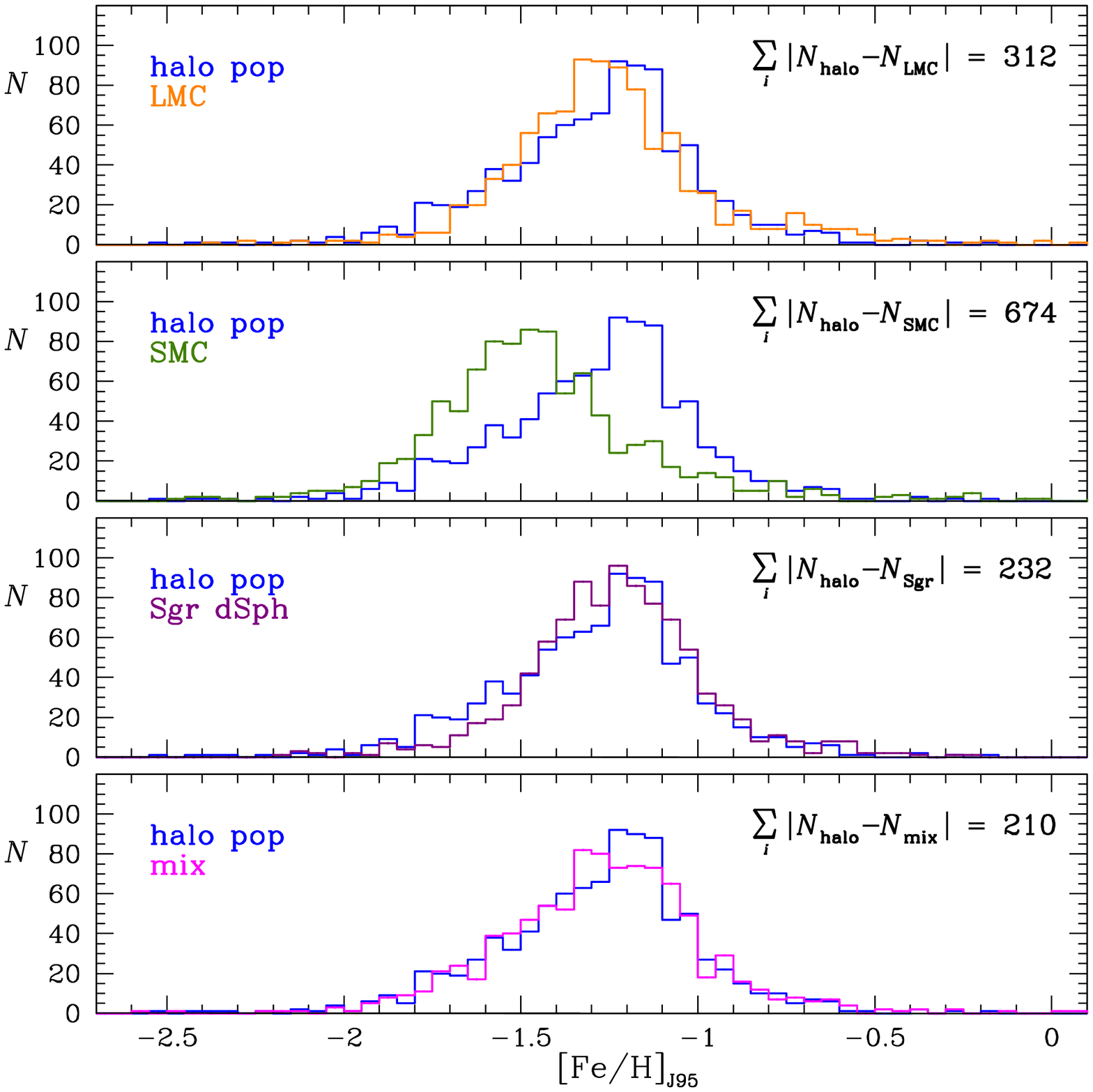}}
\FigCap{
Comparison between the metallicity distribution for the Milky Way's halo
(blue histogram in all panels) with distributions normalized to the same number
of stars for LMC, SMC, and disrupted Sgr dSph galaxy (three panels from
the top) and for a mixture of equal number of stars from these three nearby
galaxies (bottom panel). The latter distribution is the most similar to the halo
distribution. It has the lowest sum of deviations calculated over the
metallicity bins. This experiment suggests that the whole Milky Way's halo
is formed from infalling material.
}
\end{figure}


\Section{Period--amplitude diagrams}

In Figs. 6 and 7, we present period--($I$-band)amplitude diagrams (Bailey
diagrams) for the inner bulge, outer bulge, disk area, and halo.
For a transparent view, we plot only bulge variables with a small
scatter around the light curve fit ($\sigma<0.02$ mag), or stars
without the Blazhko effect. In the case of less numerous disk and halo
samples, we show all detected RRab stars. The presence of two adjacent
sequences in the Bailey diagram, A and B, was discovered in the OGLE data
for the inner bulge by Pietrukowicz \etal (2015). Here, we report that
such a double sequence continues to the outer bulge. This result indicates
that there are no major chemical differences between the inner bulge
and the outer one.

Fig.~8 shows histograms in ${\rm log} P$ for RRab stars from five Milky Way
sections counted along the sequences in the $I$-band amplitude range
0.25--0.35 mag as marked in Figs. 6 and 7. This was done the
same way as in Pietrukowicz \etal (2015). Sequence A weakens with the
increasing distance from the Galactic center. Outside the bulge, this sequence
does not exist. In the Bailey diagram, we can see that sequence B is more
blurred in the outer bulge than in the inner one. In Fig.~8, we find that
halo and disk sequences peak at slightly higher period than sequence B.
With the increasing distance from the center, the fraction of contributing
halo stars increases and widens the sequence around ${\rm log}P\approx-0.2$.

\begin{figure}[h!]
\centerline{\includegraphics[angle=0,width=99mm]{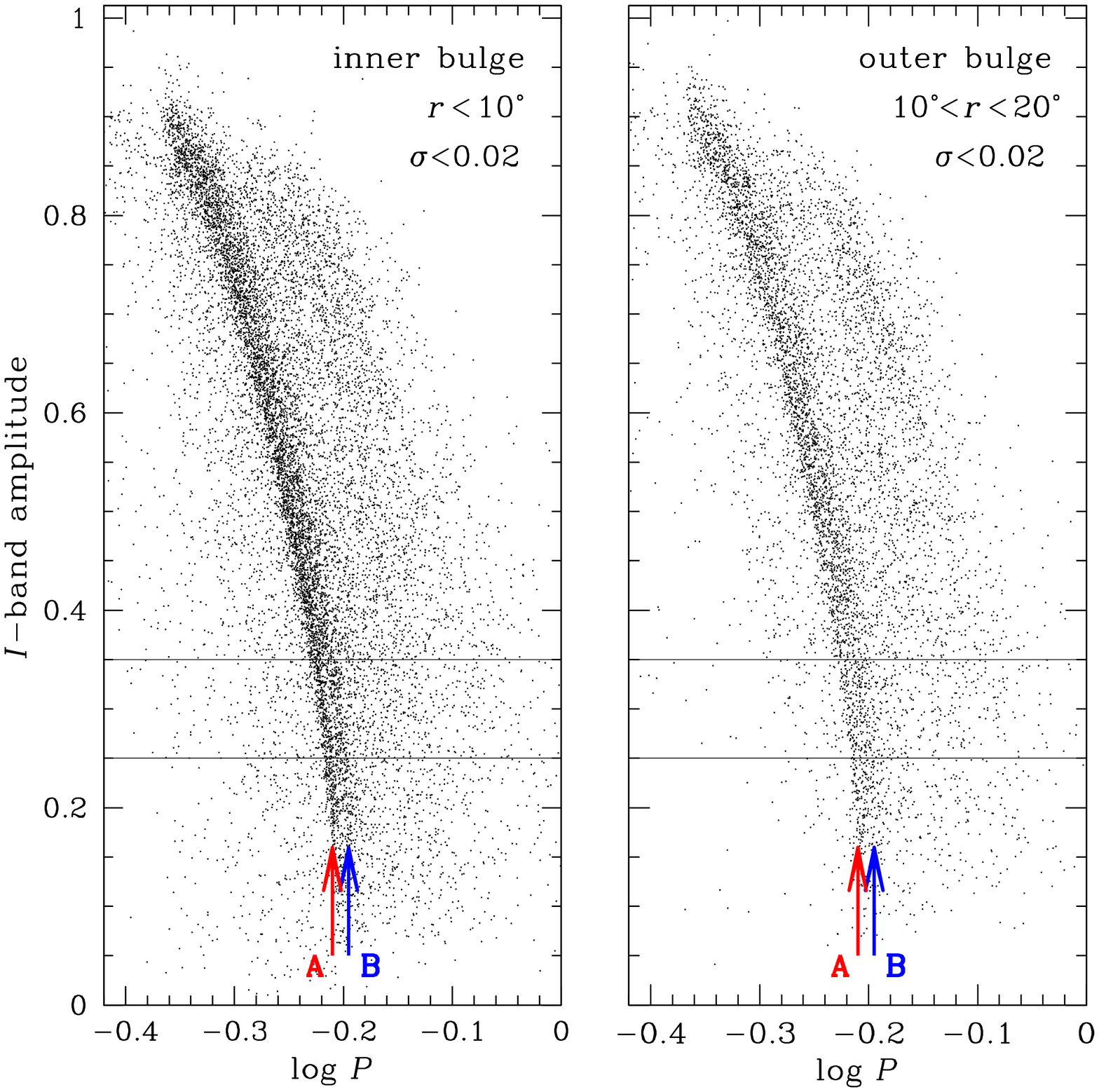}}
\FigCap{
Period-amplitude diagrams for the inner bulge (left panel) and outer bulge
area (right panel). Two major bulge sequences, A and B, are marked.
Horizontal lines denote the amplitude range used to count stars along
the sequences (see Fig.~8). Only non-Blazhko RRab stars are plotted
(with the scatter around the fit to the light curve $\sigma<0.02$ mag).
}
\end{figure}

\begin{figure}[h!]
\centerline{\includegraphics[angle=0,width=99mm]{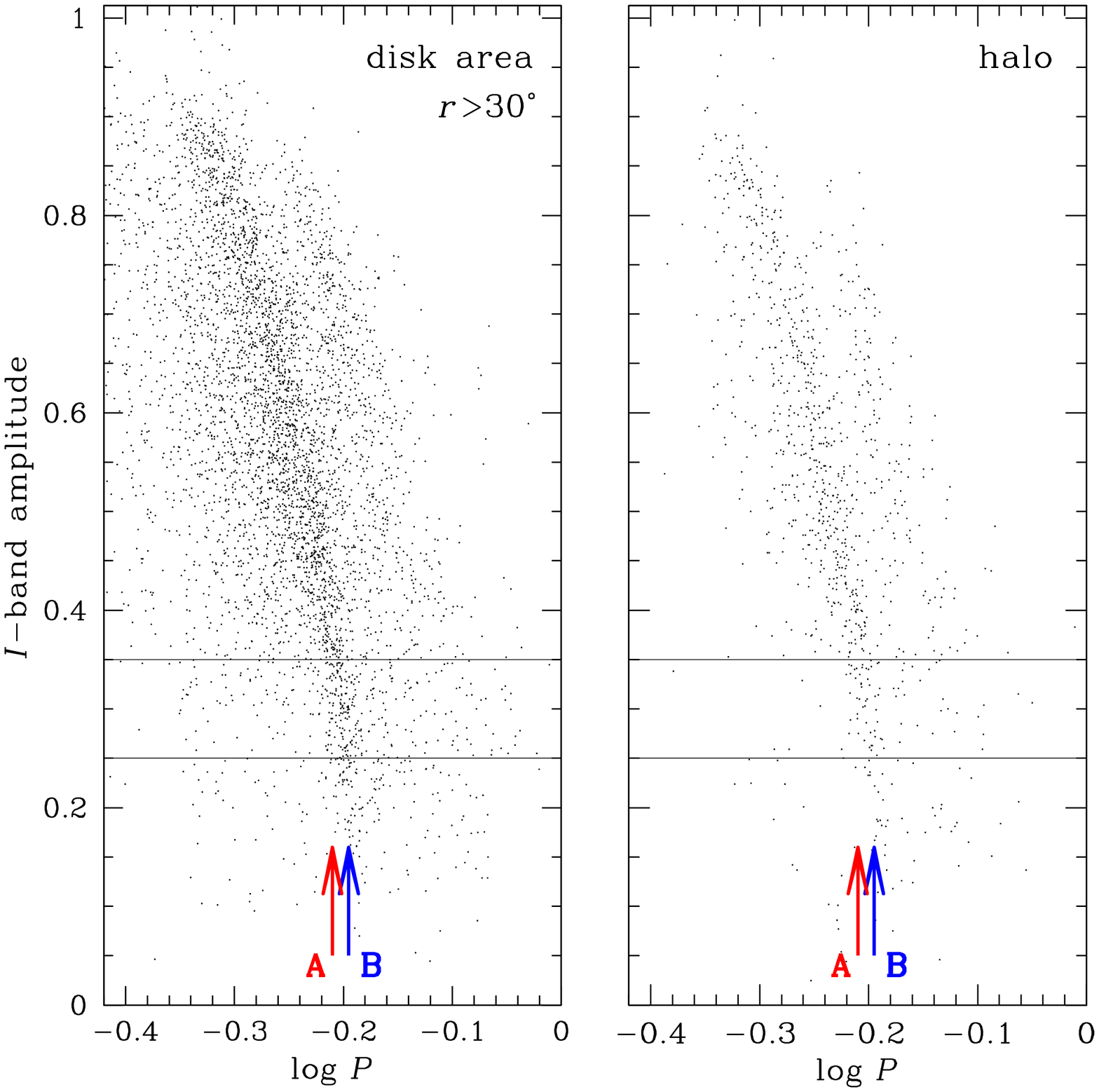}}
\FigCap{
Period-amplitude diagrams for the Galactic disk (left panel) and halo
(right panel). Locations of the bulge sequences A and B are indicated.
}
\end{figure}

\begin{figure}[htb!]
\centerline{\includegraphics[angle=0,width=80mm]{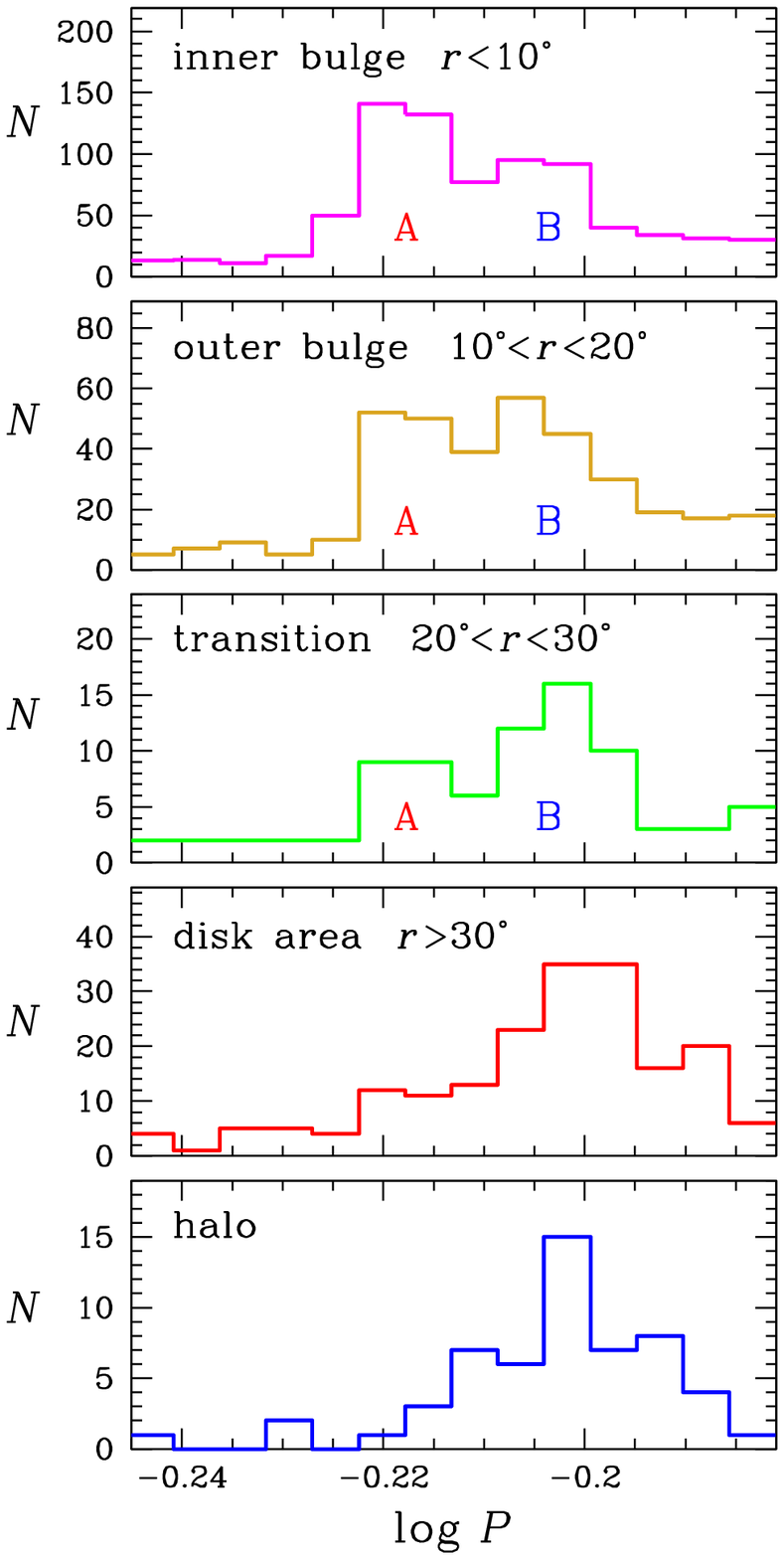}}
\FigCap{
Number histograms of RRab stars in the $I$-band amplitude range
0.25--0.35 mag in various directions of the Milky Way. Two peaks
corresponding to the sequences A and B are clearly seen in the bulge area.
}
\end{figure}

In Fig. 9, we demonstrate that the old bulge cannot be treated as an inner
extension of the halo or, more precisely, that the bulge sequence B is not
an extension of the halo Oosterhoff group I (OoI). In the period--amplitude
diagrams for the inner bulge, outer bulge, disk area, and halo, we draw curved
stripes along the major sequence. The analyzed stripes are of the same size,
limited by two fourth-order polynomials and amplitudes of 0.05 and 0.85 mag.
Metallicity profiles for stars from the major sequence of the bulge,
disk, and halo have different shapes and ranges. In particular, the bulge
and halo populations are very distinct from each other.

\begin{figure}[htb!]
\centerline{\includegraphics[angle=0,width=124mm]{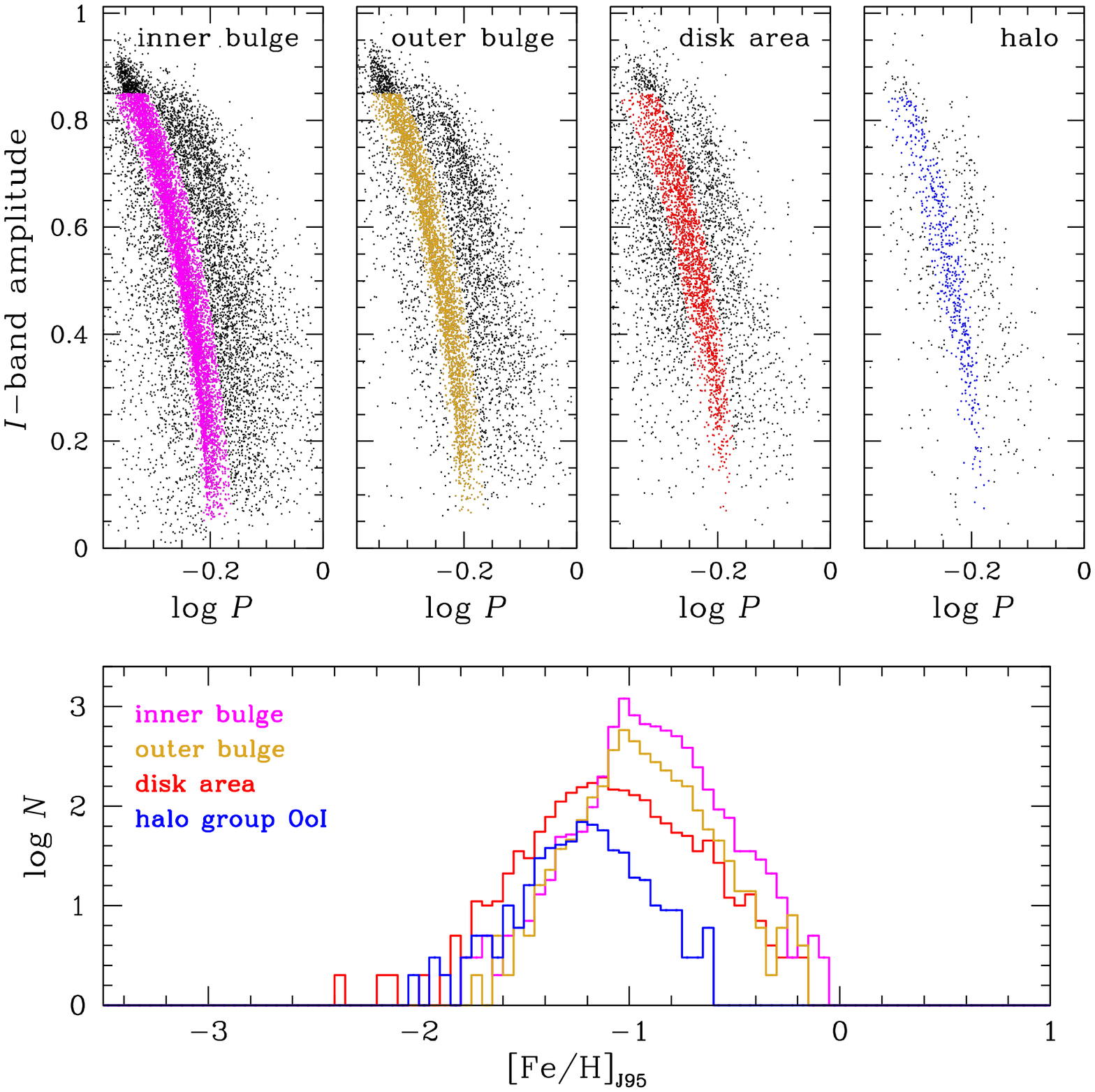}}
\FigCap{
Metallicity distributions (lower panel) for RRab stars along the major
sequence in the period-amplitude diagram in four observed directions in
the Milky Way (upper panels). The analyzed section in the Bailey diagram
is the same in all Milky Way directions. Note that the metallicity distributions
for the Galactic bulge, disk, and halo are different from each other.
}
\end{figure}


\Section{Analysis of Old Populations in the Disk Area}

RR Lyr stars in the observed disk area ($r>30\arcd$) belong to
the halo and old disk populations. It is impossible to separate
the stars based on photometry only. In Fig.~10, however, we make an attempt
to extract the old disk population metallicity profile by scaling and
subtracting the halo profile from the profile for the observed disk area.
The scaling factor was found by minimizing the difference between the
metal-poor sides of the two distributions, for ${\rm [Fe/H]_{J95}}\leq-1.2$ dex.
The result is affected by the small number statistics.
The true maximum of the disk population is at about $-0.6$ dex and the
boundary value between the disk and the halo is located at $-0.8$ dex.

\begin{figure}[htb!]
\centerline{\includegraphics[angle=0,width=124mm]{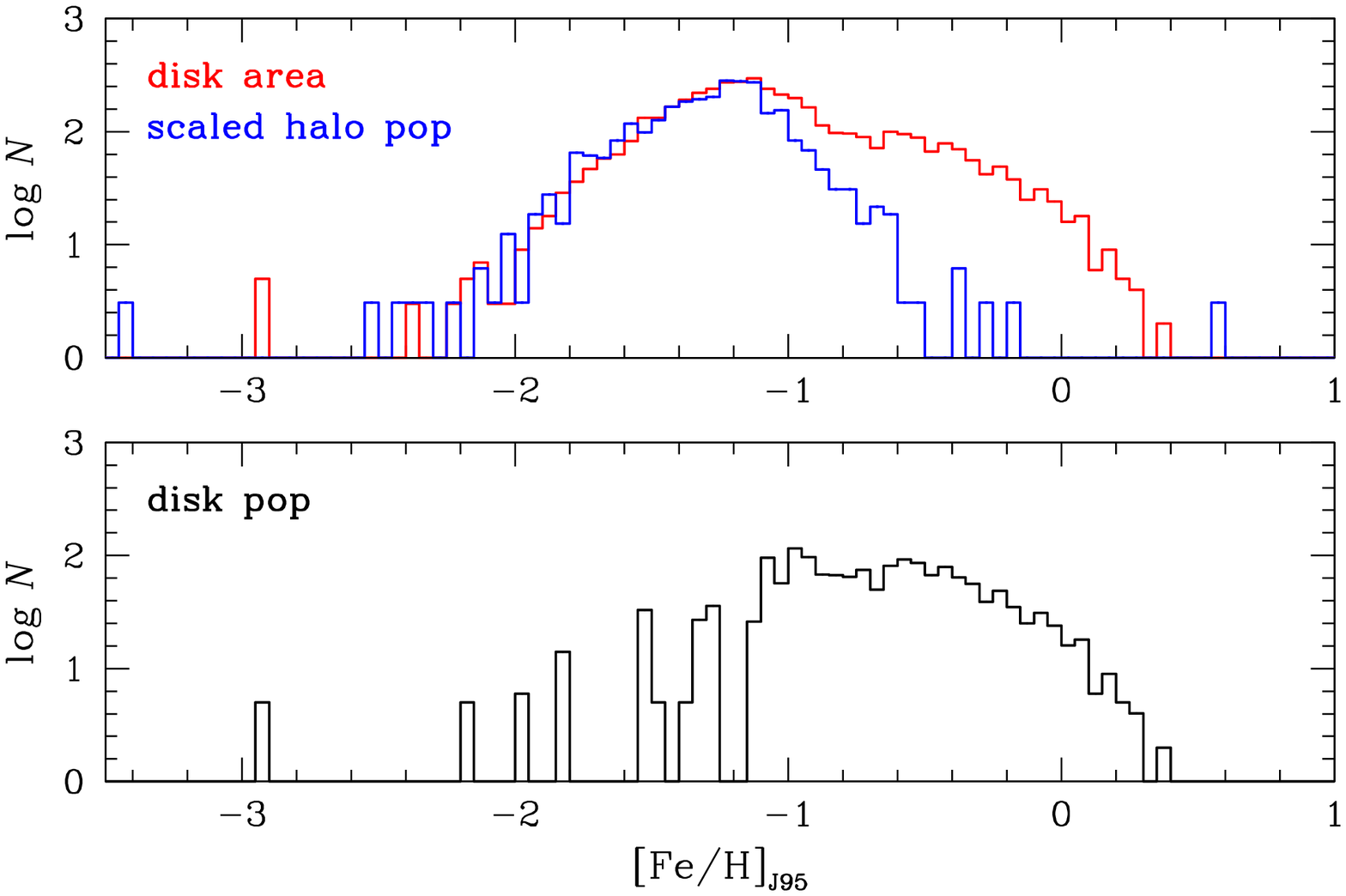}}
\FigCap{
Extraction of the metallicity distribution for the disk population
from the sample of RRab stars observed in the disk area
by subtraction of the halo population profile.
The metallicity of the disk population spreads roughly from $-1.2$
to $+0.3$ dex with the true maximum value at about $-0.6$ dex.
}
\end{figure}

In Fig.~11, we draw separate maps for 2992 metal-poor (${\rm [Fe/H]}<-0.8$ dex)
and 961 metal-rich (${\rm [Fe/H]}>-0.8$ dex) RRab stars. We present solely
stars with Galactic longitude $l<-30\arcd$ because the area closer to the
bulge is contaminated with stars from that component. In the disk area,
the fraction of metal-rich to metal-poor stars is about one-third and
it slightly decreases with the increasing angular distance from the Galactic
center (see bottom panel in Fig.~11). This result is reliable for $l>-110\arcd$,
where the observed area is symmetric with respect to the Galactic plane
and the density of stars is large enough.

Fig.~12 shows number histograms for the metal-poor and metal-rich
samples as functions of the Galactic latitude and its absolute value.
We limited the sample to the range $-30\arcd>l>-110\arcd$. There is a small
local peak at $b\approx-3\arcd$. This is rather a serendipitous arrangement
of stars emerging from clouds of dust than a real stream. The distribution
of RR Lyr stars seems to be symmetric with respect to the Galactic plane.
We do not see a structure similar to the Galactic warp in young populations
(\eg classical Cepheids, Skowron \etal 2019). The histogram
as a function of $|b|$ shows that the metal-poor and metal-rich RRab stars
have different latitudinal distributions. When the metal-rich group gets
less abundant beyond $|b|\approx3\arcd$, the number of metal-poor stars
still increases. It is evident that the metal-rich group corresponds
to the old (thick) disk and the metal-poor one belongs to the Galactic halo.

\begin{figure}[htb!]
\centerline{\includegraphics[angle=0,width=95mm]{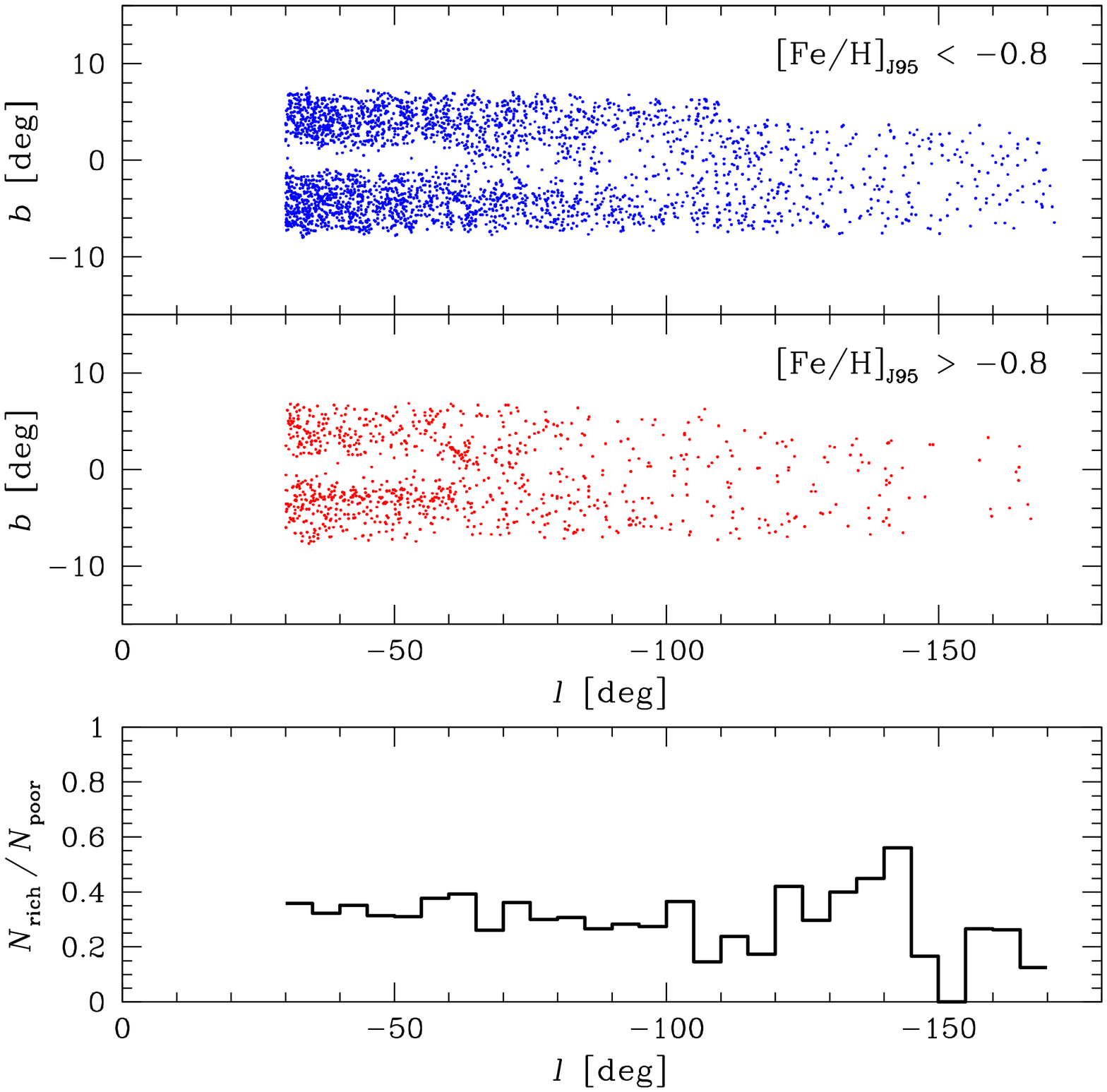}}
\FigCap{
Distribution of metal-poor (upper panel) and metal-rich RRab stars (middle panel)
detected in the disk area. Lower panel: proportion of the stars as a
function of Galactic longitude.
}
\end{figure}

\begin{figure}[htb!]
\centerline{\includegraphics[angle=0,width=95mm]{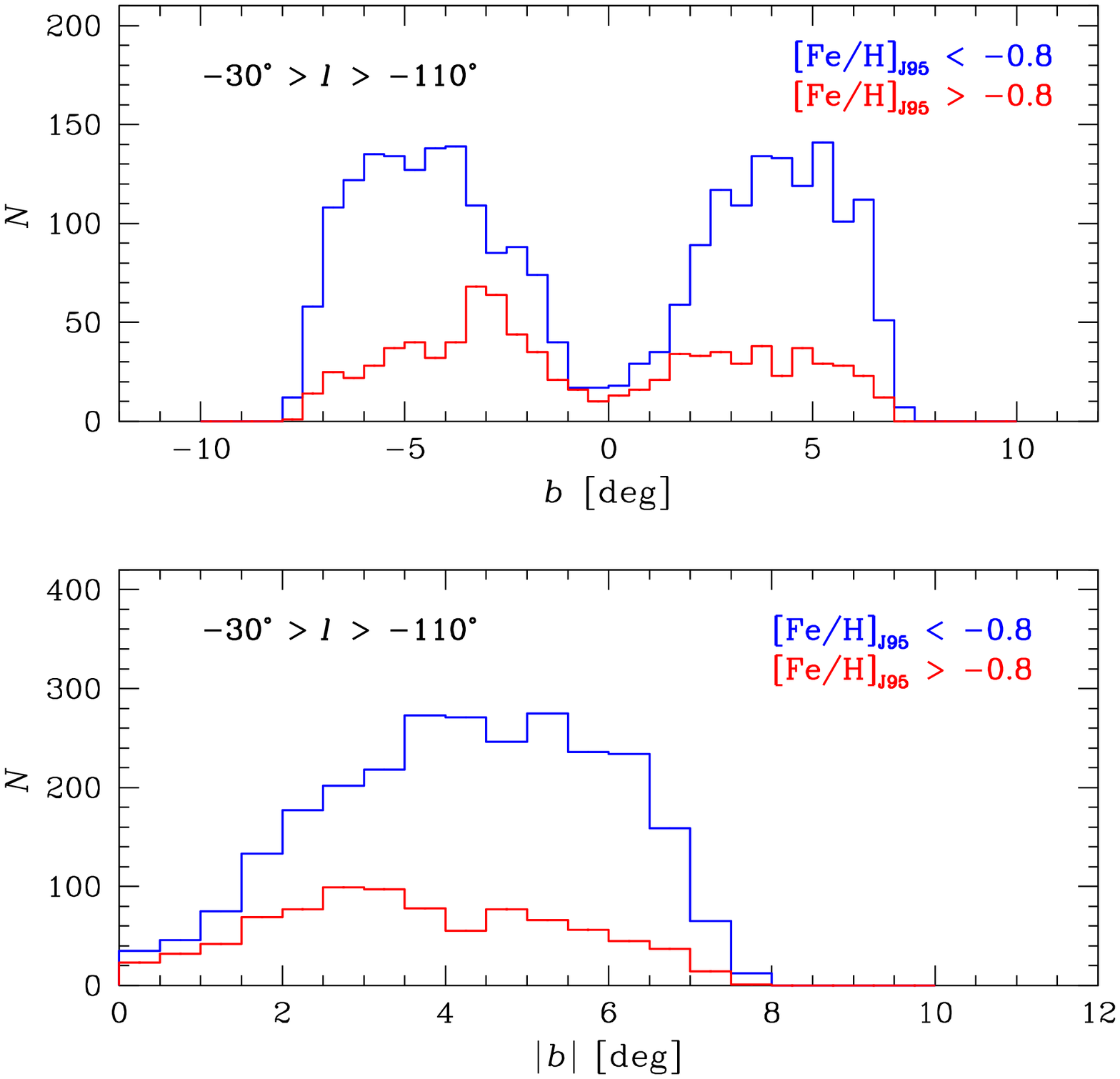}}
\FigCap{
Number histograms of metal-poor (blue) and metal-rich RRab stars (red)
along the disk as a function of $b$ (upper panel) and $|b|$ (lower panel).
Completeness of the OGLE sample drops beyond $|b|\approx6\arcd$ due to 
the boundary of the observed footprint. Note that the number of metal-rich
stars decreases closer to the Galactic plane than the metal-poor ones.
This suggests that the former stars belong to the old disk population
while the latter ones to the halo population.
}
\end{figure}

We tried to find possible relations between the Galactic coordinates,
kinematics and iron abundance of RRab stars in the disk area. We cross-matched
our disk sample with the {\it Gaia} DR2 source catalog (Gaia Collaboration \etal 2018).
This was done for 4325 stars with $|l|>30\arcd$ and derived photometric metallicities.
We adopted a matching radius of $0\zdot\arcs4$. For further investigation
only stars without duplicates and stars with {\it Renormalised Unit Weight Error}
${\rm RUWE}<1.4$ providing reliable astrometric solution were accepted.
Almost 97\% of the sources passed these criteria. In the next step,
we transformed proper motions of 4187 stars from equatorial to Galactic
system using equations given in Poleski (2013).

Fig.~13 presents distributions of median proper motions along the Galactic
longitude ($\mu_l$) and latitude ($\mu_b$) determined in seven $20\arcd$-wide
bins in the range $-30\arcd>l>-170\arcd$ and one bin for $+40\arcd>l>+30\arcd$.
The distribution of the longitudinal proper motions forms a sinusoid.
This is in agreement with the situation in which the Sun orbits the
Galactic center in a kinematically hot halo. The obtained latitudinal proper
motions are negative for the whole observed disk area with a mean value
of $-0.17\pm0.02$ mas/yr. The offset seems to result from the motion
of the Sun toward the north Galactic pole with respect to the local standard
of rest (LSR).

In Fig.~14, we show distributions of the latitudinal proper motions and their
dispersion as a function of metallicity. We limit the plot to the area
$-30\arcd>l>-110\arcd$ which is symmetric with respect to the Galactic plane.
The median value of $\mu_b$ decreases slightly with the increasing iron
abundance. In agreement with our expectations, the dispersion
$\sigma_{\mu_b}$ drops with the increasing metallicity. Surprisingly,
there is an outlying value at ${\rm [Fe/H]_{J95}}=-1.0$ dex.

\begin{figure}[htb!]
\centerline{\includegraphics[angle=0,width=95mm]{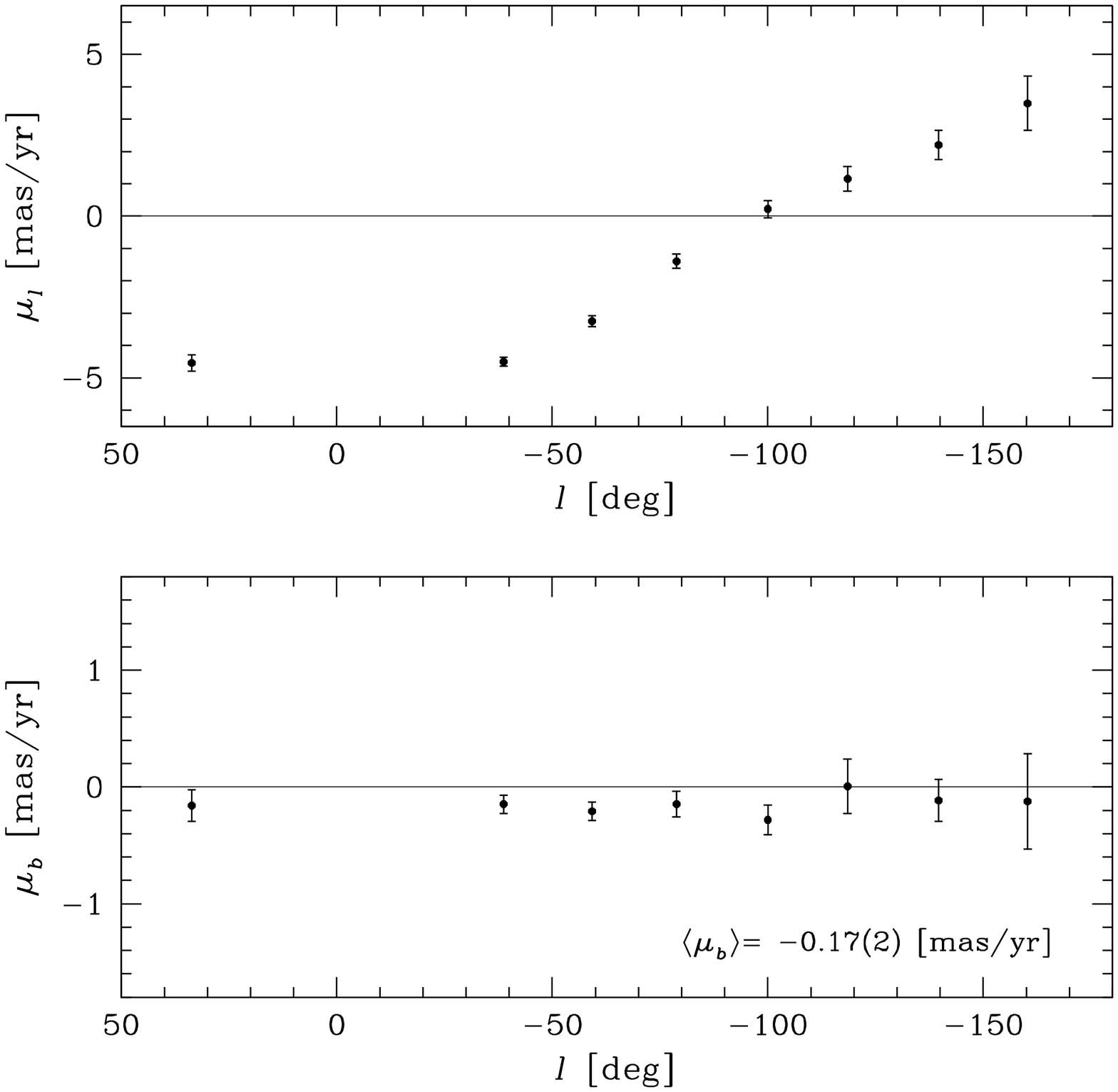}}
\FigCap{
Longitudinal (upper panel) and latitudinal proper motions (lower panel)
of the disk RRab stars in the function of Galactic longitude. The observed
sinusoidal trend in $\mu_l$ is consistent with the Sun orbiting the Milky
Way's center in a kinematically hot halo. The value of $\mu_b$ is negative
along practically the whole observed disk. This likely stems from the
vertical motion of the Sun with respect to the local standard of rest. 
}
\end{figure}

\begin{figure}[htb!]
\centerline{\includegraphics[angle=0,width=95mm]{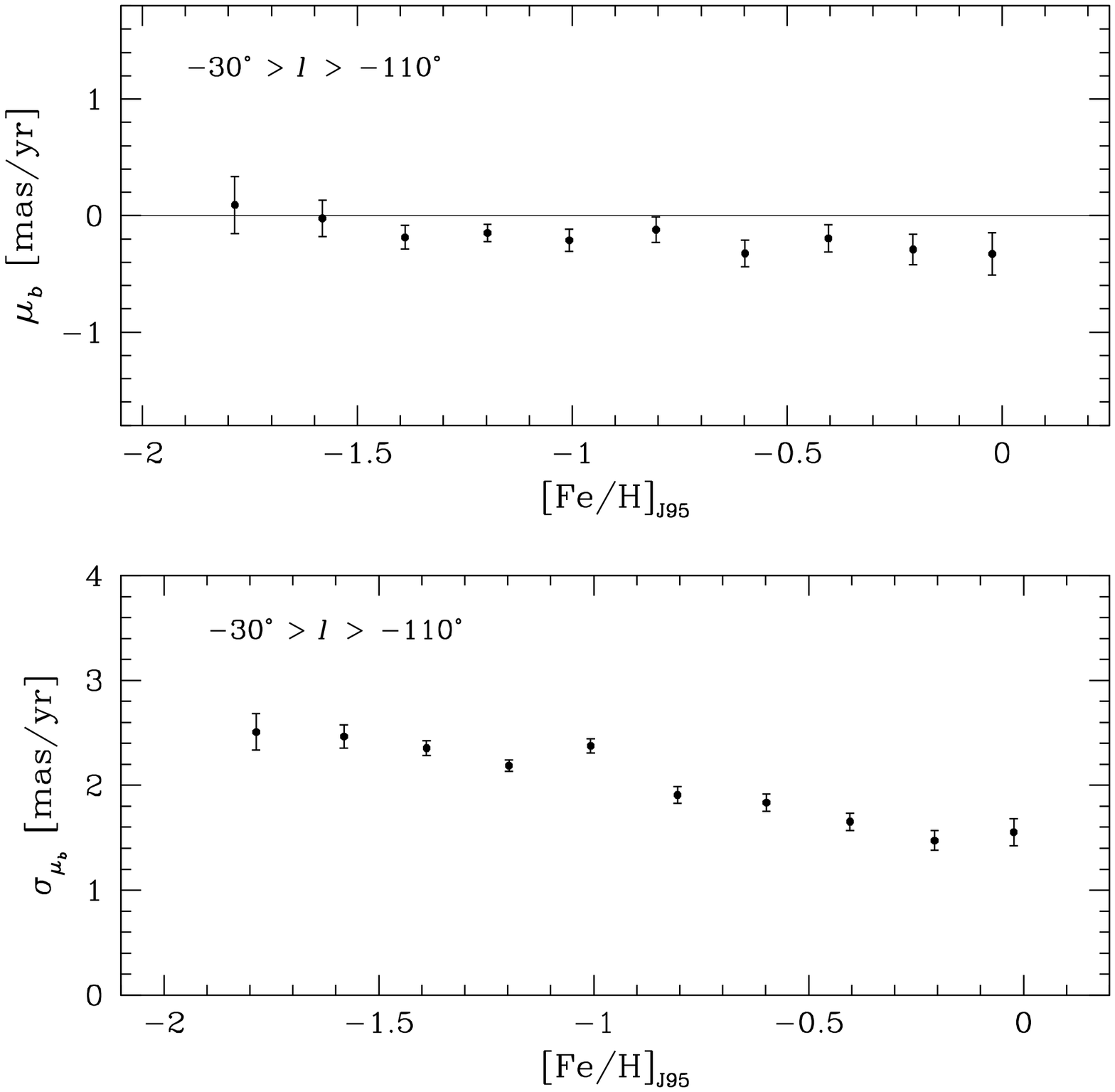}}
\FigCap{
Latitudinal proper motion (upper panel) and its dispersion (lower panel)
as a function of metallicity for RRab stars from the Galactic disk area
with $-30\arcd>l>-110\arcd$. Note that $\sigma_{\mu_b}$ decreases with the
increasing metal content except for a significant bump around
${\rm [Fe/H]_{J95}}=-1.0$ dex.
}
\end{figure}


\Section{Conclusion}

We used the largest available homogeneous collection of Galactic RRab
stars, the OGLE collection, to describe ancient stellar populations of the
Milky Way based on photometric metallicity. From the metallicity distributions
we concluded that there are solely three old components: the Galactic
halo, Galactic bulge, and Galactic disk. A comparison of RR Lyr stars from
the inner and outer bulge indicates that the entire old bulge is rather
chemically homogeneous. The distributions of the inner and outer parts
have a very similar shape and both peak at ${\rm [Fe/H]_{J95}}\approx-1.0$ dex.
We reported the presence of two adjacent sequences in the period--amplitude
diagram, or populations A and B, also in the outer bulge. From the shape of
the metallicity distributions we inferred that halo stars are found inside the bulge.
About one-third of the RR Lyr stars in the bulge area, within $20\arcd$ from
the center, belong to the Galactic halo. This fraction drops to a quarter
within $10\arcd$. Our result is consistent with expectations and the kinematic
study of nearly 2800 bulge RR Lyr stars conducted by Kunder \etal (2020),
who found that 25\% of the variables are halo interlopers. Our experiment with
mixing LMC, SMC, and Sgr dSph RR Lyr populations in equal proportions may
suggest that the entire Milky Way's halo is formed from infalling dwarf
galaxies, as we obtained a very similar metallicity distribution of the
mixed populations to the one of the halo.
The distribution for RR Lyr variables from the disk area has two peaks:
one at ${\rm [Fe/H]_{J95}}\approx-1.2$ dex corresponding to the halo
component and the other one at ${\rm [Fe/H]_{J95}}\approx-0.6$ dex
corresponding to the disk component. The two components have different
vertical number distributions. We thus confirmed that the disk population
represented by RR Lyr stars exist, which has been implied based on several
times smaller samples (\eg Kinemuchi \etal 2006, Mateu and Vivas 2018,
Prudil \etal 2020). The spatial distribution of RR Lyr stars is symmetric
with respect to the Galactic plane and no warp in the old disk population
is seen. There is no evidence for the presence of multiple (more than two)
components in the metallicity distribution for RRab stars in the disk area
as found from near-infrared observations by D\'ek\'any \etal (2018).
On-going spectroscopic and astrometric surveys (such as BRAVA-RR, APOGEE,
{\it Gaia}) will provide more precise information on chemical composition
and kinematics of the OGLE RR Lyr stars, which should allow for more
detailed conclusions on the Milky Way's old populations.


\Acknow{
We thank OGLE observers for their contribution to the collection of
the photometric data over the years. The OGLE project has received
funding from the National Science Centre, Poland, grant MAESTRO
2014/14/A/ST9/00121 to A.U. This work has been supported by the
National Science Centre, Poland, grants OPUS 2016/23/B/ST9/00655
to P.P. and MAESTRO 2016/22/A/ST9/00009 to I.S.
In this research, we used data from the European Space Agency (ESA)
mission {\it Gaia}, processed by the {\it Gaia} Data Processing and
Analysis Consortium (DPAC). Funding for the DPAC has been provided by
national institutions, in particular the institutions participating
in the {\it Gaia} Multilateral Agreement.}


\end{document}